# The solitary wave in advanced nuclear energy system

Jin Feng Huang *

Department of Nuclear science and Engineering, East China University of Technology, Nanchang 330000, China

**Abstract**

The solitary wave naturally arises in many areas of mathematical physics, including in nonlinear optics, plasma physics, quantum field theory and fluid mechanics. For advanced nuclear energy system, travelling wave reactor or CANDLE reactor, has been proposed for couples of years. However, the analytical solution has not been discovered. In this study, according to the perspective of solitary wave, the analytical solution of this advanced nuclear energy system is demonstrated through coupling one group neutron diffusion equation with burnup equation. The tanh-function method is applied to solve that nonlinear partial differential equation. The relationship between velocity of solitary wave, wave amplitude or neutron flux and the evolution of nuclide is revealed by analytical method.

1. **Introduction**

The solitary wave in non-linear system had been discovered by John Scott in Russell 1844 (J. S. Russell, 1844). Solitary waves are particle-like waves that own to a balance between nonlinear and dispersive effects. Solitary wave propagates without any temporal evolution in shape or size when it moves at constant speed and conserves amplitude, shape, and velocity. Solitary wave naturally arises in many areas of mathematical physics (E. Infeld and G. Rowlands, 1990), including in nonlinear optics (L. F. Mollenauer et al., 1980), plasma physics (N. J. Zabusky, 1965), quantum field theory and fluid mechanics. The classical example of an equation yielding solitary wave solutions is the Korteweg-de Vries equation (D. J. Korteweg and G. Vries, 1895), which is model of waves on shallow water surfaces. The solitary wave also can be observed in advanced nuclear energy system (Walter Seifritz, 1998), travelling wave reactor or CANDLE (Constant Axial shape of Neutron flux, nuclide number densities and power shape During Life of Energy producing reactor) reactor (Sekimoto et al., 2001).

* Corresponding author：Jin Feng Huang, fjndhjf@ecut.edu.cn

A concept of completely automated nuclear reactor for long-term operation was proposed by Teller (Teller et al., 1996). This reactor core which comprises of ignition region and breeding region is quite different from the conventional reactor. The neutrons leaking from ignition region are captured by fertile fuel which is subsequently converted the fissile fuel in breeding region. The breeding region is adopted with thorium material but 50% of the heavy metal nuclides are depleted at the end of the lifetime of the reactor. The self-stabilizing criticality waves in such reactor were presented by Hugo Van Dam (H. Dam, 2000). The analytical model with reactivity feedback was illustrated through introducing a parabolic burnup function which is the most simple form, and the ignition condition for a criticality wave was provided. Base on the same neutronic model with reactivity feedback, a two-group diffusion model coupled with simplified burn-up equations is investigated (Xue Nong Chen et al., 2007) for a one-dimensional burn-up drift wave problem.

The feasibility of creating self-organizing breeding/burning waves was analyzed by S. Fomin (S. Fomin et al., 2005; S. Fomin et al., 2008). At the starting phase, the requirements for wave initiation and evolution in space and time were discussed through coupling diffusion transient model with burnup equations. Breed-and-burn strategy in a fast reactor with optimized starter fuel was carried out by J. Huang (Jin Feng Huang et al., 2015). The optimization of the starter fuel was performed to reduce the initial positive excess reactivity swing and to flatten the power distribution, and the results show that elaborating the ignition region is effective to reduce fuel enrichment and improve the operation performance during the starting phase.

A particular class of travelling wave reactor, called the CANDLE reactor, was illustrated firstly by H. Sekimoto (Sekimoto et al., 2001). CANDLE reactor could be fed by natural or depleted uranium, and could show a variety of attractive characteristics (Sekimoto, H. and Miyashita, S., 2006; Sekimoto, H. and Nagata A., 2009; Sekimoto, H. and Nakayama Sinsuke, 2014; Y. Ohoka and H. Sekimoto, 2004). The equilibrium state of the CANDLE burnup was discussed through solving numerically neutronic diffusion equations coupled with burnup equations. In the equilibrium state, distributions of nuclide densities, neutron flux, and power density are remained constant shapes and the same constant speed for constant power operation. The equilibrium state of the CANDLE can also be considered as solitary wave formed by neutron flux starting to propagate steadily.

For the solitary wave, only by retaining the balance between the nonlinear and dispersive effects can the solitary wave be remained. The dispersive effects represent in leakage term in the nonlinear partial differential equation (PDE). Consequently, dealing with the nonlinear effects caused by neutron fission and absorption in medium is the key point to produce and propagate the solitary wave. In this study, the nonlinear PDE was constructed through coupling neutron diffusion equation with burnup equation. Usually, nonlinear PDE is difficult to attain the analytical solution. However, if the nonlinear term in above PDE was expanded by Taylor's series, then the tanh-function method could be applied to solve this PDE. The necessary boundary condition and simplification were adopted to get the analytical solution. The results show that the neutron fluxes, neutron fluences and the evolution of nuclide density can be represented as the form of solitary wave. Even though numerical solutions were widely developed to solve the PDE, the analytical solutions are still desired since the numerical approaches have to be recalculated for every set of parameters, but analytical approaches allow someone to have all answers.

## 2. Nonlinear partial differential equation
### 2.1 Burnup equations

For the heavy nuclides I, the general burnup equations could be written as:

$$\frac{\partial N_I}{\partial t} = -\sigma_a^I \Phi N_I + \sigma_c^{I-1} \Phi N_{I-1} - \gamma_I N_I + \gamma_{I' \to I} N_{I'} \quad (1)$$

Where $N_I$, $\Phi$, $\sigma_a^I$, $\sigma_c^I$, $\gamma_I$, $I'$ denote atomic density of nuclide $I$, neutron flux, microscopic absorption cross section of nuclide $I$, microscopic capture cross section of nuclide $I$, radioactive decay constant of nuclide $I$, mother nuclide of $I$, respectively. In an actual situation, for $^{238}$U-$^{240}$Pu conversion chains, the production of $^{239}$Pu (e.g., by the neutron capture of $^{238}$U) and decay processes should be considered, but they are omitted here for the sake of simplicity. Therefore, $^{238}$U→$^{239}$U→$^{239}$Np→$^{239}$Pu→$^{240}$Pu conversion chains can be simplified as $^{238}$U→$^{239}$Pu→$^{240}$Pu (or fission productions). Even through this simplification would loss some accuracy, but it would not change the physical inherence and make the issue more easily. For $^{238}$U, the burnup equation can be expressed as:

$$\frac{\partial N_8}{\partial t} = -\sigma_{a8} \Phi N_8 \quad (2)$$

The solution of this differential equation is as followed:

$$N_8 = N_{8,0} e^{-\sigma_{a8} \Psi} \quad (3)$$

$$\Psi = \int_{t_0}^{t} \Phi dt$$

Where $N_{8,0}$ denotes the initial atomic density of $^{238}$U. Similarly, for $^{239}$Pu, burnable poisons (BP) and fission productions (FP) the burnup equation can be expressed as:

$$\frac{\partial N_9}{\partial t} = -\sigma_{a9} \Phi N_9 + \sigma_{c8} \Phi N_8 \quad (4)$$

$$\frac{\partial N_{BP}}{\partial t} = -\sigma_{aBP} \Phi N_{BP} \quad (5)$$

$$\frac{\partial N_{FP}}{\partial t} = \sum_i \sigma_{fi} N_i \Phi \tag{6}$$

The solutions of these differential equations are as followed:

$$N_9 = N_{9,0} e^{-\sigma_{a9}\Psi} + N_{8,0} \frac{\sigma_{c8}}{\sigma_{a9} - \sigma_{a8}} [e^{-\sigma_{a8}\Psi} - e^{-\sigma_{a9}\Psi}] \tag{7}$$

$$N_{BP} = N_{BP,0} e^{-\sigma_{aBP}\Psi} \tag{8}$$

$$N_{FP} = \sum_i \sigma_{fi} N_i \Psi \tag{9}$$

Where $N_{9,0}$ denotes the initial atomic density of $^{239}$Pu, and nuclide $i$ could be chosen as $^{238}$U and $^{239}$Pu for fission reaction. The value of $N_{9,0}$ can be equal to zero for CANDLE reactor in breeding region since the production of $^{239}$Pu is only through the neutron capture of $^{238}$U. One of commons for these solutions should be emphasized that they can be expressed as exponential functions even if take account into $^{240}$Pu and $^{241}$Pu in the burnup chain. The exponential functions still can be expressed as the Taylor's series, which provides one way of analytically solving the nonlinear PDE.

### 2.2 Neutron diffusion theory

The one-group neutron diffusion model plays an important role in reactor theory even through it is sufficiently simple. This simple diffusion model coupled with burnup equation also sufficiently realistic to reveal the producing and propagating of the solitary wave. The one-group neutron diffusion (Duderstadt and Hamilton, 1976) can be expressed here,

$$D \frac{\partial^2 \Phi}{\partial x^2} + (\upsilon \Sigma_f - \Sigma_a) \Phi = \frac{1}{v} \frac{\partial \Phi}{\partial t} \tag{10}$$

Where $D$ denotes the neutron diffusion coefficient, $\Sigma_a$ the macroscopic absorption cross section, $\upsilon$ average neutron number per fission, $\Sigma_f$ macroscopic fission cross section, $v$ neutron speed. Furthermore,

$$\Sigma_a = \sum_i N_i \sigma_{ai}, \ \upsilon \Sigma_f = \sum_i \upsilon_i N_i \sigma_{fi}, \ D = \frac{1}{3\Sigma_{tr}}$$

$N_i$ could be substituted to the solution of burnup equations, such as, $N_8$, $N_9$, etc. $\Sigma_{tr}$ is the macroscopic transport cross-section. In term of Eq.(7)~ Eq.(9), all the expressions of $N_i$ contact with the factor $e^{-\sigma \Psi}$. Hence, the second term of Eq.(10) in the left hand, $(\upsilon \Sigma_f - \Sigma_a)$ multiplied $\Phi$, is a nonlinear term in PDE. Only these nonlinear effects cancelling out dispersive effects, the solitary wave can propagate over large distances but without dissipation. The $(\upsilon \Sigma_f - \Sigma_a)$ is equal to:

$$(v\Sigma_f - \Sigma_a) = (v_9 N_9 \sigma_{f9} + v_8 N_8 \sigma_{f8} + v_0 N_0 \sigma_{f0}) - (N_8 \sigma_{a8} + N_9 \sigma_{a9} + N_0 \sigma_{a0} + N_{FP} \sigma_{aFP} + N_{BP} \sigma_{aBP}) \quad (11)$$

All the $N_i$ ($i$=238, 239, 240, BP) are the function of exponential form and can be expanded as Taylor's series. The $^{239}$Pu fissions caused by fast neutrons are dominant compared with $^{238}$U fissions and $^{240}$Pu fissions in fast neutron spectrum. Therefore, $^{238}$U fissions and $^{240}$Pu fissions are neglected. All the terms in Eq. (11) taken into account for calculation are fine and can improve the accuracy only need to repeat expansion behavior, but such behaviors increase the complication of solving nonlinear PDE. The appendix C shows the $N_9 \sigma_{a9}$ and $N_{FP} \sigma_{aFP}$ was taken into account for calculation in details. For the sake of simplicity, only the $v_9 N_9 \sigma_{f9}$, $N_8 \sigma_{a8}$ are remained to reveal the propagating of solitary wave clearly. Consequently, the nonlinear PDE can be rewritten:

$$D \frac{\partial^2 \Phi}{\partial x^2} + (v_9 N_9 \sigma_{f9} - N_8 \sigma_{a8})\Phi = \frac{1}{v} \frac{\partial \Phi}{\partial t} \quad (12)$$

$$F(\Psi) \equiv (v\Sigma_f - \Sigma_a) = N_{8,0} \left[ \left( v_9 \frac{\sigma_{c8} \sigma_{f9}}{\sigma_{a9} - \sigma_{a8}} - \sigma_{a8} \right) e^{-\sigma_{a8} \Psi} - v_9 \frac{\sigma_{c8} \sigma_{f9}}{\sigma_{a9} - \sigma_{a8}} e^{-\sigma_{a9} \Psi} \right] \quad (13)$$

Where $e^{-\sigma_{a8}\Psi}$ and $e^{-\sigma_{a9}\Psi}$ can be expanded by second order Taylor's series and here omit the higher order terms. This approximation could be available since the magnitude of $\sigma_{a9}$ is several barns and of $\Psi$ is $10^{20}$ cm$^{-2}$ (H. Dam, 2000) in a fast spectrum type CANDLE reactor. For a thermal spectrum type CANDLE reactor, this approximation could also be available since $\sigma_a \Psi$ is still small even through the $\sigma_a$ increases but $\Psi$ decreases. Therefore, the value of $\sigma_a \Psi$ is small that $e^{-\sigma_a \Psi}$ can be expanded by second order Taylor's series.

$$e^{-\sigma_{a8}\Psi} = 1 + (-\sigma_{a8}\Psi) + \frac{1}{2}(-\sigma_{a8}\Psi)^2$$

$$e^{-\sigma_{a9}\Psi} = 1 + (-\sigma_{a9}\Psi) + \frac{1}{2}(-\sigma_{a9}\Psi)^2$$

Surely, adopting the second order Taylor's series would provide the higher accuracy for the smaller $\sigma_a \Psi$.

3. **Analytical solution and discussion**

3.1 **Analytical solution**

In this section, the processes of pursuing analytical solution of nonlinear PDE are illustrated in detail. Usually, the nonlinear PDE is difficult to have analytical solution. It is discovered that this type PDE would have analytical solution by using the tanh-function method (Huibin Lan and Kelin Wang, 1989) if the nonlinear terms are expanded by Taylor's series. Although the analytical solution methods such as the inverse scattering transform (M. Ablowitz et al., 1974), homogeneous balance method (E. Fan and H. Zhang, 1998), the (G'/G)-expansion method (A. Bekir, 2008) were developed. However, those analytical methods are in most cases difficult to handle and require a thorough knowledge of its properties and possibilities. The tanh-function method is a common powerful method for solving nonlinear equations, and play an important role in problems where reaction, dispersive effects, diffusion and/or convection (W. Malfliet, 2004).

$$D\frac{\partial^2 \Phi}{\partial x^2} + F(\Psi)\Phi = \frac{1}{v}\frac{\partial \Phi}{\partial t} \qquad (14)$$

**Step 1.** Using the travelling wave transformation,

$$\Phi(x,t) = \Phi(c(x - ut)) = \Phi(\eta), (u > 0, \ c > 0) \qquad (15)$$

Where u is phase speed of wave, $c$ is wave numbers and it represents the wave with a characteristic width $L=c^{-1}$, and η means coordinate.

$$\frac{1}{v}\frac{\partial \Phi(x,t)}{\partial t} = -\frac{cu}{v}\frac{d\Phi(\eta)}{d\eta} \qquad (16)$$

$$\frac{\partial^2 \Phi}{\partial x^2} = c^2 \frac{d^2 \Phi(\eta)}{d\eta^2}$$

Therefore, Eq.(14) can be transformed into the ordinary differential equation (ODE):

$$c^2 D \frac{d^2 \Phi(\eta)}{d\eta^2} + \frac{cu}{v}\frac{d\Phi(\eta)}{d\eta} + F(\Psi)\Phi(\eta) = 0 \qquad (17)$$

**Step 2.** Assume that the solution of Eq.(17) has the form,

$$\Phi = \sum_{j=0}^{n} a_j T^j, \quad T \equiv tanh(\eta) \qquad (18)$$

Where $a_j$ is constant coefficient, $T$ denotes hyperbolic tangent function. The integer *n* is depended on the balance between the highest order derivatives and the nonlinear terms.

**Step 3.** The highest order of derivatives terms $\frac{d^2\Phi}{d\eta^2}$ is $n + 2$, and the highest order of nonlinear terms $F(\Psi)\Phi$ is $2(n-1) + n$. The proof would be represented in Appendix A. Consequently, we obtain $n = 2$ and then Φ could be expressed:
$$\Phi = a_0 + a_1 T + a_2 T^2$$

Now the boundary conditions ($\Phi(\pm\infty) = 0$) are taken into account this expression. Even though the neutron flux distribution of

CANDLE reactor is not always limited in narrow region, but in this research, only the practical case in which the CANDLE reactor is constrained in narrow region was investigated. For a solitary wave, the function value must be close to a constant number if coordinate is away enough. In the CANDLE reactor, solitary wave which is limited in narrow region other than distributes entire region, the constant number should be zero since the neutrons have the finite mean free path in medium. The research (Sekimoto et al., 2001) shows that if the core height is set at 8 m then the characteristic of CANDLE reactor could be obtained. If the core height is less than 8 m or the value of tanh-function is not small enough, the boundary condition would not be suitable well. Applying the boundary, according to Fig. 1, two expressions $\Phi(+\infty) = a_0 + a_1 + a_2 = 0$ and $\Phi(-\infty) = a_0 - a_1 + a_2 = 0$ can be got. Subtracting or adding two expressions, the coefficients would be solved. Therefore, the coefficient $a_1$ should be zero, not only this, but also $a_0$ should be equal to $-a_2$ then the boundary conditions could be yielded. Consequently,

$$\Phi = -a_2 + a_2 T^2, (a_2 < 0) \tag{19}$$

The coefficient $a_2$ should be negative real numbers since the neutron flux should be positive and finite real numbers.

The neutron fluence $\Psi$ can be obtained by the integral of $\Phi$,

$$\Psi = \int \Phi(\eta) dt = -\frac{a_2}{cu}(1-T) \tag{20}$$

The appendix B shows the proof. Substitute Eq.(20) into Eq.(13), the expression can be written,

$$F(\Psi) = N_{8,0}\left((C_8 - C_9) + \frac{-a_2(1-T)}{cu}(C_9\sigma_{a9} - C_8\sigma_{a8}) + \frac{1}{2}\left(\frac{-a_2(1-T)}{cu}\right)^2(C_8(\sigma_{a8})^2 - C_9(\sigma_{a9})^2)\right) \tag{21}$$

$$C_8 = \left(v_9 \frac{\sigma_{c8}\sigma_{f9}}{\sigma_{a9} - \sigma_{a8}} - \sigma_{a8}\right)$$

$$C_9 = v_9 \frac{\sigma_{c8}\sigma_{f9}}{\sigma_{a9} - \sigma_{a8}} = C_8 + \sigma_{a8}$$

**Step 4**. Substituting Eq.(19), Eq.(20) and Eq.(21) into Eq.(17) and collecting all the terms with the same power $T^i$, i = 0,1,2,3,4.

$$c^2(1-T^2)[-2T(a_1 + 2a_2T) + 2a_2(1-T^2)] + \frac{cu}{Dv}(1-T^2)(a_1 + 2a_2T) + F(\Psi)\frac{\Phi}{D} = 0 \tag{22}$$

$T^3$ coefficient: $\left(-\frac{2cua_2}{Dv} - \frac{a_2^2 C_8 \sigma_{a8} N_{8,0}}{cDu} - \frac{a_2^3 C_8 \sigma_{a8}^2 N_{8,0}}{c^2Du^2} + \frac{a_2^2 C_9 \sigma_{a9} N_{8,0}}{cDu} + \frac{a_2^3 C_9 \sigma_{a9}^2 N_{8,0}}{c^2Du^2}\right) = 0, (a_1 = 0)$

$T^4$ coefficient: $\left(6c^2 a_2 + \dfrac{a_2^3 C_8 \sigma_{a8}^2 N_{8,0}}{2c^2 D u^2} - \dfrac{a_2^3 C_9 \sigma_{a9}^2 N_{8,0}}{2c^2 D u^2}\right) = 0$

Equating each coefficient of this polynomial to zero yields a set of algebraic equations for $a_n$ (n = 0, 1, 2). Due to $a_1$ determined by boundary condition, $T^3$ and $T^4$ coefficients of these polynomials were selected. Solving the equation system, we can construct a variety of analytical solutions for Eq. (22). Solving the $T^4$ and $T^3$ coefficient of these polynomials, Eq. (23), Eq. (24) and Eq. (25) can be obtained separately,

$$a_2 = -\dfrac{2\sqrt{3D}uc^2}{\sqrt{N_{8,0}}\sqrt{-C_8\sigma_{a8}^2 + C_9\sigma_{a9}^2}} \quad (a_2 < 0) \tag{23}$$

$$a_2 = \dfrac{2\sqrt{3D}uc^2}{\sqrt{N_{8,0}}\sqrt{-C_8\sigma_{a8}^2 + C_9\sigma_{a9}^2}} \quad (a_2 > 0) \tag{24}$$

$$a_2 = \dfrac{2c^2 u(-u + 6cDv)}{v(C_8\sigma_{a8} - C_9\sigma_{a9})N_{8,0}} \tag{25}$$

The Eq.(24) should be rejected because $a_2$ should be less than zero to yield the boundary condition due to $C_9 > C_8$ and $\sigma_{a9} > \sigma_{a8}$. The solution of the Eq.(25) sometimes shows a positive real numbers and could not always yield the boundary condition ($a_2 < 0$) since it depend on the value of numerator, therefore, it could be rejected.

Consequently, the neutron flux $\Phi$ has the analytical solution,

$$\Phi(x,t) = \dfrac{2\sqrt{3D}uc^2}{\sqrt{N_{8,0}}\sqrt{-C_8\sigma_{a8}^2 + C_9\sigma_{a9}^2}}\left(1 - \tanh^2(c(x - ut))\right) \tag{26}$$

If the $N_9\sigma_{a9}$ and $N_{FP}\sigma_{aFP}$ were taken into account for calculation, Eq.(26) would be updated as:

$$\Phi(x,t) = \dfrac{2\sqrt{3D}c^2 u}{\sqrt{N_{8,0}}\sqrt{-C_8\sigma_{a8}^2 + C_9\sigma_{a9}^2 - 2C_{fp}\sigma_{a8} + 2C_{fp}\sigma_{a9}}}\left(1 - \tanh^2(c(x - ut))\right)$$

Where $C_8$, $C_9$, $C_{fp}$ would be redefined as Appendix C. The results present that the differences are slightly if the $N_9\sigma_{a9}$ and $N_{FP}\sigma_{aFP}$ were taken into account but would not change the trends.

For a given medium, the Eq. (23) demonstrates that wave velocity u is lineally proportional to the wave amplitude $a_2$ as a result of the nonlinear character of the wave, and Eq. (23) also shows that wave velocity u is proportional to $N_{8,0}$ since $D$ is proportional to $1/N_{8,0}$ if neutron flux is fixed. These conclusions are consistent with Hugo Van Dam (H. Dam, 2000) and H. Sekimoto (Sekimoto et al., 2001). The wave velocity depended on the wave amplitude which is kept with constant is the characteristic for a soliton (Lamb, 1980). What's more, Eq. (23) also reveals the relationship between wave amplitude and microscopic cross section of nuclide $^{238}$U and $^{239}$Pu. The Eq. (23) explicitly shows that the neutron flux is independent of neutron speed $v$, however, both the diffusion coefficient $D$ and microscopic cross section are depended on neutron speed implicitly. The parameters of solitary wave can be adjusted easily to meet the requirements in term of Eq. (23) or Eq. (26). It should be mentioned that a critical reactor can operate at any flux level, hence, the magnitude of flux is depended on the power level of the core. The free coefficient $c^2$ is still unknown and could be determined by the thermal power output $P$ to normalize neutron flux:

$$P = \int_{-\infty}^{\infty} Q_f \Sigma_{f9} \Phi(\eta) dx = \frac{4\sqrt{3D} u c \Sigma_{f9}}{Q_f \sqrt{N_{8,0}} \sqrt{-C_8 \sigma_{a8}^2 + C_9 \sigma_{a9}^2 - 2C_{fp}\sigma_{a8} + 2C_{fp}\sigma_{a9}}}$$

Where $Q_f$ is the energy produced per fission (taken as 200 MeV).

The neutron fluence can be represented:

$$\Psi(\eta) = \frac{-a_2}{cu}(1 - tanh(\eta)) = \frac{2\sqrt{3D}c}{\sqrt{N_{8,0}}\sqrt{-C_8\sigma_{a8}^2 + C_9\sigma_{a9}^2 - 2C_{fp}\sigma_{a8} + 2C_{fp}\sigma_{a9}}}(1 - tanh(\eta)) \qquad (27)$$

### 3.2 Discussion
### 3.2.1 The profile of neutron flux and neutron fluence

It should be mentioned that the Eq.(23) and Eq.(26) will provide us with the profile of neutron flux since the free coefficient $c$ is still unknown. The magnitude of neutron flux or nuclide density should be determined by the thermal power output of the reactor core. For a given medium, Eq. (26) exhibits that the profile of solitary wave would be lanky if neutron flux is increased due to the wave number $c$ increased. Eq. (26) also implies that the amplitude of neutron flux is linearly proportional to the square root of diffusion coefficient, but is inversely proportional to initial $^{238}$U density since diffusion coefficient $D$ is inversely proportional to initial $^{238}$U density and other materials. This illustrates that reducing the fuel density increases the amplitude of neutron flux, but the total reaction rate should keep the same if the power is constant, therefore, the changing makes the solitary wave lanky. The larger amplitude of solitary wave is not desirable since it increases the power peaking factor notably. Oppositely, the lower fuel density would broaden the distribution of power. The one of concerns for CANDLE reactor is that the high power peaking factor takes great challenge for reactor safety. This study provides insight into solving this issue.

Table 1 shows the selected parameters taken from H. Sekimoto (Sekimoto et al., 2001) and Xue-Nong Chen (Xue-Nong Chen et al., 2012) to apply and verify the equations. In Table 1, even though the diffusion coefficient $D$ contains some part of information about medium, but in this simplified model, the burnable poisons and coolant would be ignored. The $\Phi_{max}$ is the maximum neutron flux under a given condition, and here is applied to one dimension rather than multi-dimensions. According to the provided parameters, the wave number $c$ which was considered as the only free parameter could be determined in term of Eq. (26), then the profile of neutron flux, neutron fluence and fuel density, from Fig.2 to Fig.6, could be draw according to the analytical solutions. The Fig.2 shows that the profile of neutron flux is bell-shaped solitary wave which is drifting as time going on.

Eq. (27) indicates that the neutron fluence is inversely proportional to initial $^{238}$U density, which denotes that the lower fuel density would have higher neutron fluence. Eq. (27) also displays that the maximum neutron fluence is none business of wave speed and depended on the characteristics of medium.

Table 1. The selected one-group parameters of fast neutron spectrum

| Parameter | Value |
|---|---|
| $D$ | 1.470 cm |
| $\Phi_{max}$ | $3.0 \times 10^{15}$ cm$^{-2}$s$^{-1}$ |
| $N_{8,0}$ | $0.01221 \times 10^{24}$ cm$^{-3}$ |
| $\upsilon\sigma_f, \sigma_f, \sigma_a, \sigma_c, \sigma_{tr}$ | ~barn |
| $^{238}$U | 0.142, 0.051, 0.404, 0.352, 8.181. (barn) |
| $^{239}$Pu | 5.878, 2.007, 2.481, 0.474, 8.593. (barn) |
| Fission Productions | 0, 0, 0.4973, 0.4973, 11.92. (barn) |
| u | $1.1 \times 10^{-7}$ cm/s |

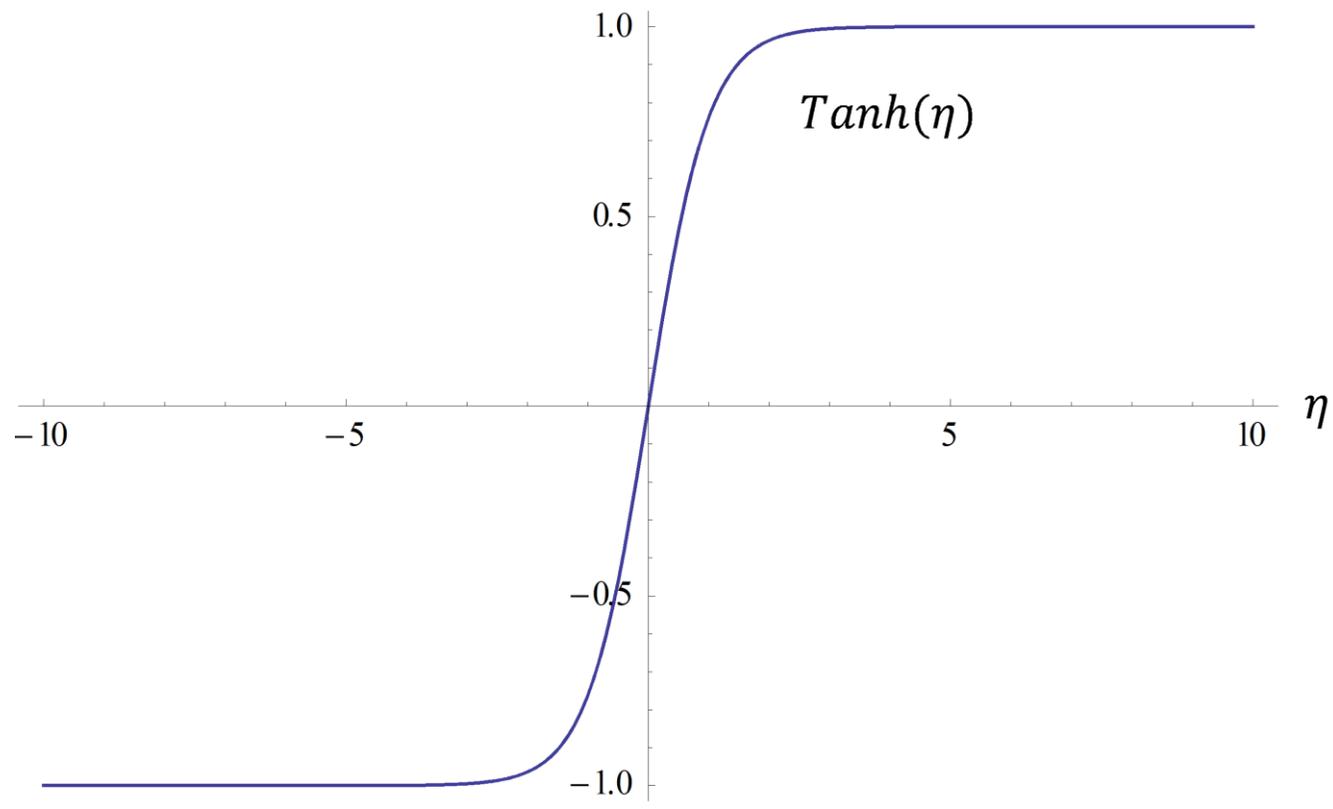

Fig. 1. The profile of Tanh function

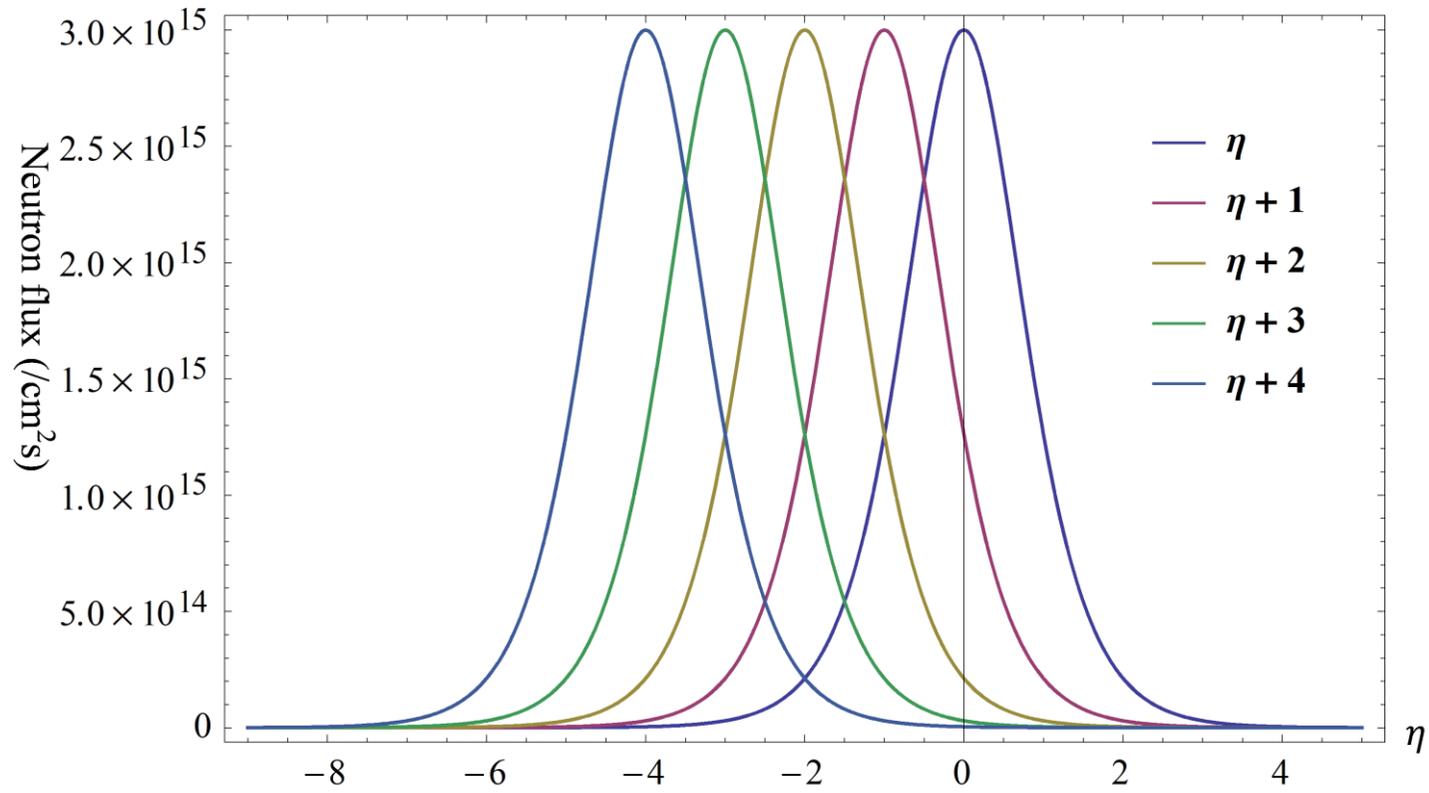

Fig. 2. The profile of neutron flux is bell-shaped solitary wave

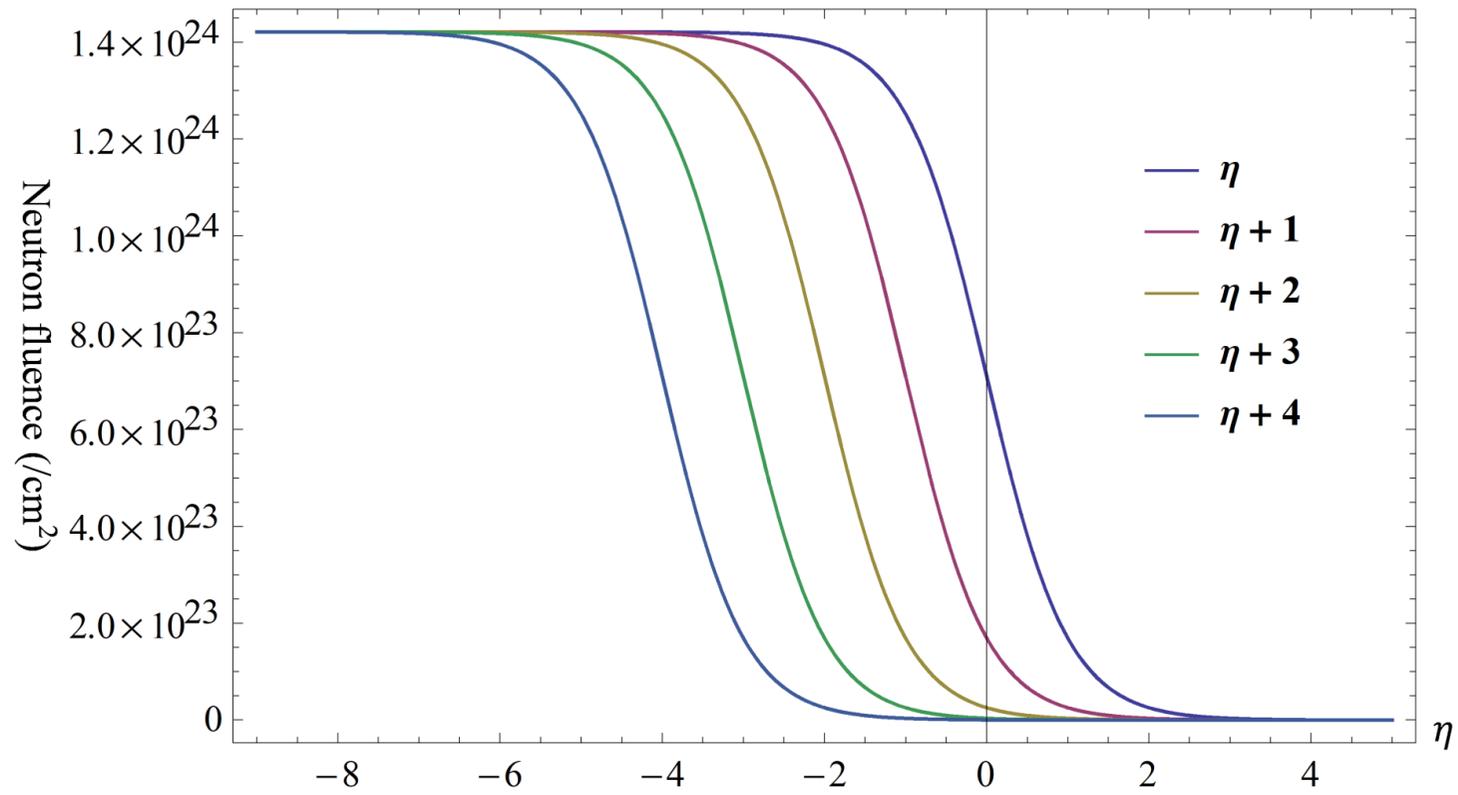

Fig. 3. The profile of neutron fluence is an anti-kink solitary wave

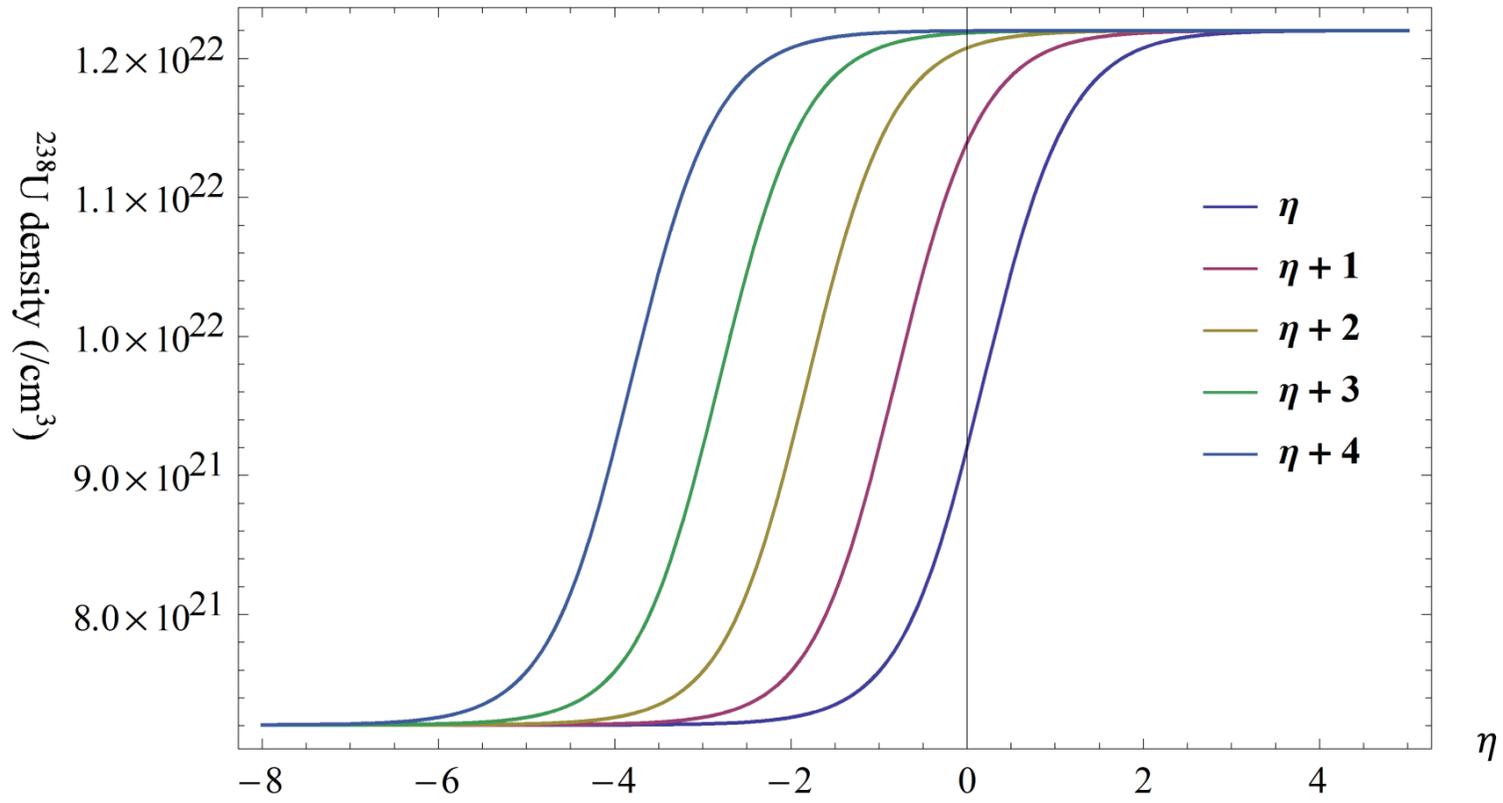

Fig. 4. The evolution of $^{238}U$ is as kink solitary wave

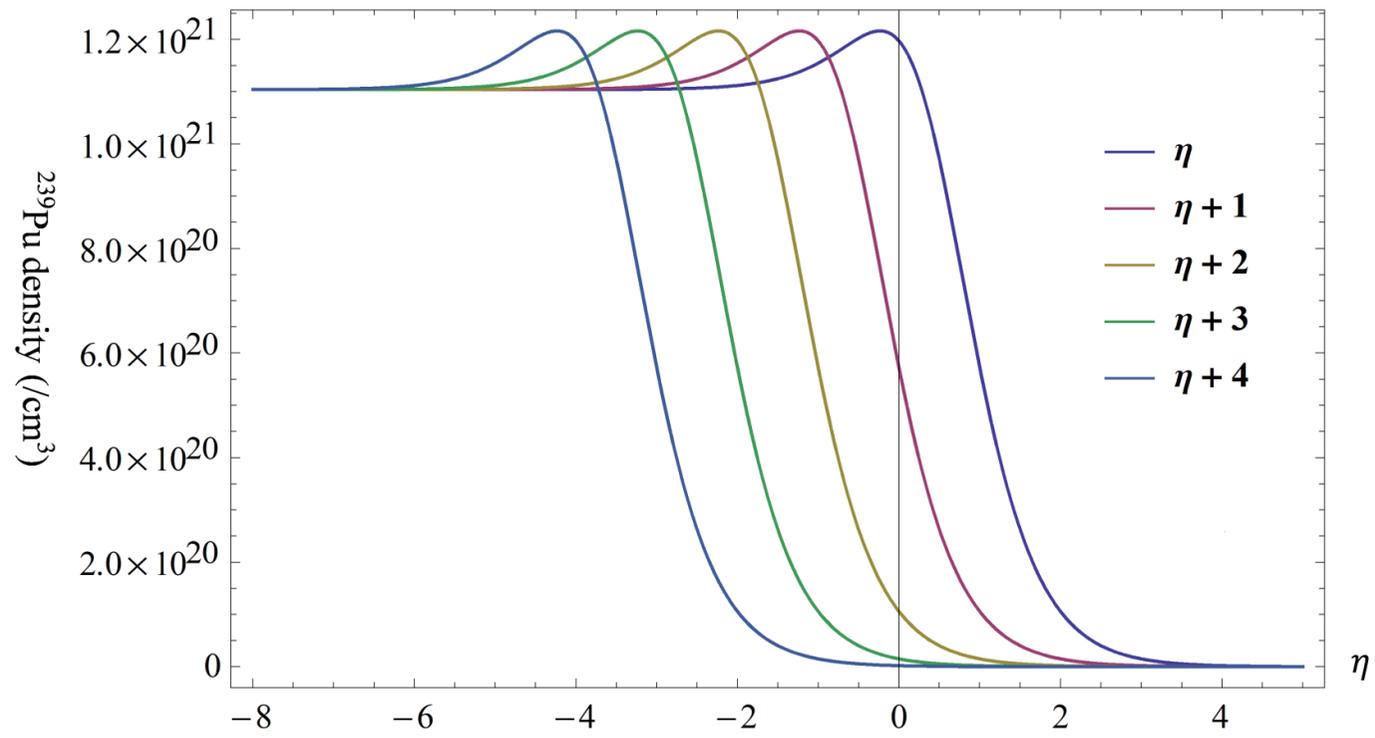

Fig. 5. The evolution of $^{239}$Pu density is as one new type solitary wave

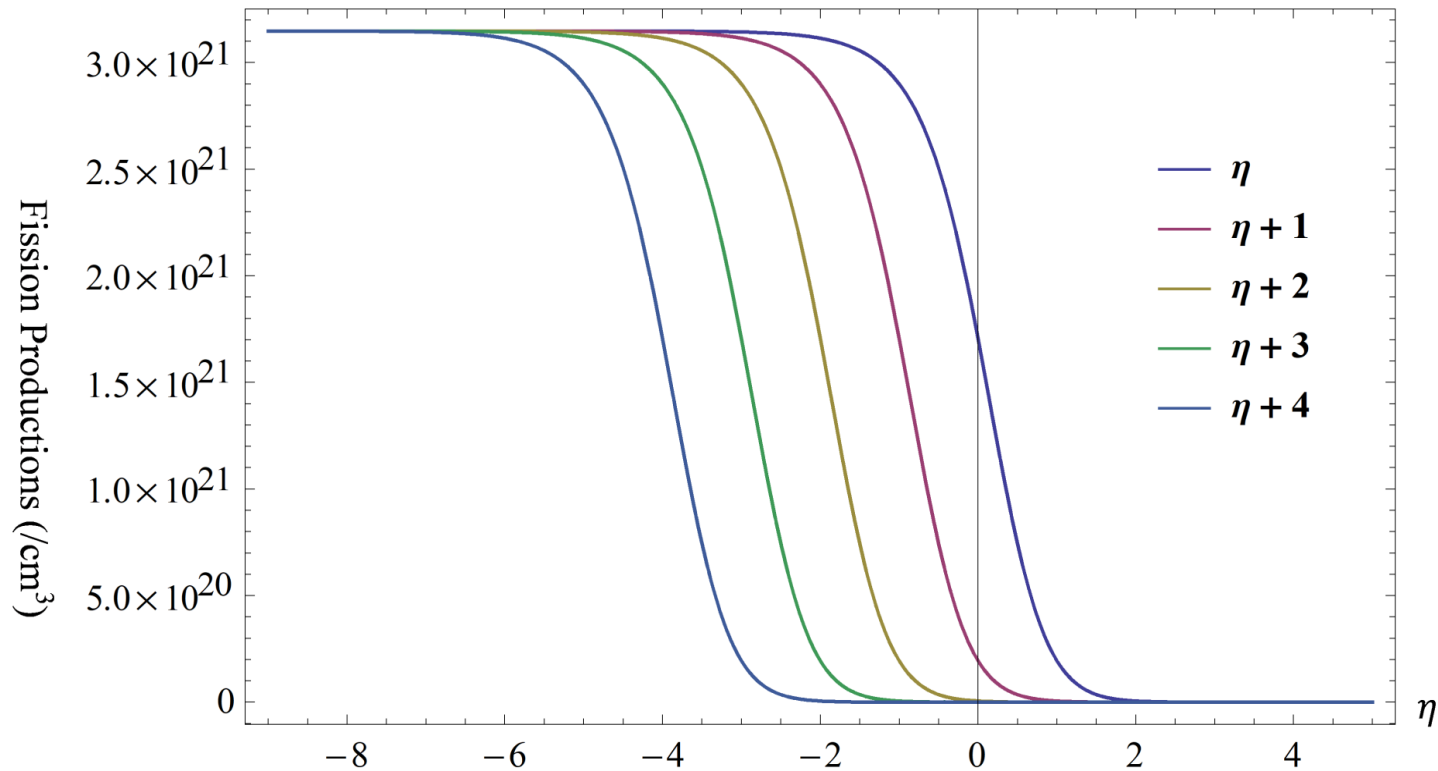

Fig. 6. The profile of fission productions is an anti-kink solitary wave

### 3.2.2 The profile of Fuel burnup and the evolution of nuclides

One of the notable merits for CANDLE reactor is that the fuel burnup can be as high as 400 Gwd/t only loading with depleted uranium or natural uranium in breeding region. The fuel burnup is linearly proportional to neutron fluence. Therefore, the fuel burnup should have same profile with Fig. 3. The trend of burnup solitary wave profile coincides well with the previous study (Jin Feng Huang et al., 2015) which is performed by Monte-Carlo method coupled ORIGEN burnup code. On the other hand, the very high fuel burnup also takes great challenge due to the fuel and cladding radiation damages for the nuclear engineering. This analytical solution of solitary wave point out that the fuel burnup can be reduced through increasing the initial fuel density due to increasing neutron fluence. The evolutions of nuclides play a vital role in maintaining solitary wave. The solutions of burnup equations reveal such evolutions. Applying the Taylor's series, the nuclide $^{238}$U can be written,

$$N_8 = N_{8,0}(1 - \Psi\sigma_{a8} + \frac{1}{2}\Psi^2\sigma_{a8}^2) \tag{28}$$

Substitute Eq.(20) into Eq.(28),

$$N_8 = N_{8,0}\left[1 - \left(\frac{\sigma_{a8}a_2}{cu} - \frac{\sigma_{a8}a_2}{cu}tanh(\eta)\right) + \frac{1}{2}\left(\frac{\sigma_{a8}a_2}{cu} - \frac{\sigma_{a8}a_2}{cu}tanh(\eta)\right)^2\right] \tag{29}$$

Similarly, the evolution of $^{239}$Pu can be expressed as,

$$N_9 = N_{8,0}\frac{\sigma_{c8}}{\sigma_{a9} - \sigma_{a8}}\left[exp\left(\frac{-a_2\sigma_{a8}}{cu} - \frac{-a_2\sigma_{a8}}{cu}tanh(\eta)\right) - exp\left(\frac{-a_2\sigma_{a9}}{cu} - \frac{-a_2\sigma_{a9}}{cu}tanh(\eta)\right)\right] \tag{30}$$

The evolutions of $^{238}$U and $^{239}$Pu profiles are displayed as Fig. 4 and Fig. 5. The trends of evolutions of $^{239}$Pu are also matched well with the previous studies (Walter Seifritz, 1998; Jin Feng Huang et al., 2015; Xue-Nong Chen et al., 2012). The $^{238}$U density is depended on microscopic absorption cross section and neutron fluence. The Fig. 4 demonstrates that consumed percentage of $^{238}$U reaches about 37% to transform into $^{239}$Pu or fission directly. The evolution of $^{239}$Pu, breeding from $^{238}$U, shows that there is a peak value toward the propagating direction. Comparing the Fig. 4 with Fig. 5, the stored density of $^{239}$Pu is approximately less ten times than the feeding $^{238}$U. In term of Fig.2 and Fig. 5, it can be found that the location of maximum neutron flux is not coincided and advanced with location of maximum density of $^{239}$Pu. Fig. 6 shows the evolution of the fission productions, and its behavior is similar to neutron fluence.

## 4. Conclusions

The analytical solutions of advanced nuclear energy system, travelling wave reactor or CANDLE reactor, is represent through coupling one-dimensional, one-group neutron diffusion equation with burnup equation, and the tanh-function method is employed to solve this nonlinear partial differential equation. In order to obtain the analytical solutions, some necessary simplifications are adopted. The analytical solutions of neutron fluxes, neutron fluences and the evolution of nuclide density can be obtained, and be represented as the different solitary waves. Finally, the parameters are selected to apply and verify the analytical solutions. The results demonstrate that the neutron flux is lineally proportional to wave velocity but inversely proportional to fuel density and microscopic absorption cross section. The profile of neutron flux is a bell-shaped solitary wave modified by wave velocity, fuel density and microscopic absorption cross section. In additional, the neutron fluence, the fuel density of $^{238}$U and $^{238}$Pu show as solitary wave. The maximum neutron fluence is independent of wave velocity and depended on the property of medium. The analytical solutions provide important insights into the physical phenomena of advanced nuclear energy system clearly. In the future, the analytical solutions of multi-dimensional with thermal-hydraulic feedbacks will be investigated to extend the scenarios.

## Acknowledgements

This work was supported by the Ignition Research Funds for doctor in East China University of technology (No. 110-1410000736). The authors are also grateful to the China Scholarship Council (CSC No. 201706310058), education department project in Jiangxi province of China (No. GJJ180402) and Special high-level talent research start-up (No.DHBK2017132 and No.110-1410000994) for their support.

**Appendix A. The parameter *n* would be determined by balancing the linear terms of highest order with the nonlinear terms.**

For the hyperbolic tangent function $tanh(\eta)$, it has the following formula:

$$\int tanh^n(\eta) d\eta = -\frac{tanh^{n-1}(\eta)}{n-1} + \int tanh^{n-2}(\eta) d\eta$$

Because of $\Phi = \sum_{j=0}^{n} a_j tanh^j(\eta)$, so the highest order of $\Phi$ is $n$. The highest order of $\Psi$ is named $O(\Psi)$,

$$O(\Psi) = O\left(\int \Phi dt\right) = O\left(-\frac{1}{u}\int tanh^n(\eta) d\eta\right) = n - 1$$

Consequently, $O(\Psi^2) = 2(n-1)$. After the $F(\Psi)$ expanded by second order Taylor series, we can obtain $O(F(\Psi)) = 2(n-1)$

Therefore, the highest order of nonlinear terms $F(\Psi)\Phi$ is $2(n-1)+n$. The highest order of derivatives term $\frac{d^2\Phi}{d\eta^2}$ is $n+2$. It can be explained as followed:

$T \equiv tanh(\eta)$

$$\frac{d\Phi}{d\eta} = (1-T^2)\frac{d\Phi}{dT}$$

$$\frac{d^2\Phi}{d\eta^2} = (1-T^2)\left(-2T\frac{d\Phi}{dT} + (1-T^2)\frac{d^2\Phi}{dT^2}\right)$$

Where

$\Phi = \sum_{j=0}^{n} a_j T^j$, that is $O(\Phi) = n$. Thus it could be obtained:

$O\left(\frac{d^2\Phi}{d\eta^2}\right) = n+2$.

The highest order of nonlinear terms and the highest order of derivatives term should keep balance, thus $2(n-1) + n = n+2$, consequently, $n$ should be equal to 2.

**Appendix B. Calculating the neutron fluence.**

$$\Psi = \int \Phi(\eta')dt = \int_{-\infty}^{t'}(-a_2 + a_2(tanh(\eta))^2)dt = \frac{-a_2}{-cu}\left[\int_\infty^\eta (1-(tanh(\eta))^2)d\eta'\right] = \frac{a_2}{cu}\left[\int_\infty^\eta (Sech^2(\eta'))d\eta'\right] = \frac{-a_2}{cu}(1-tanh(\eta)) \quad (A1)$$

**Appendix C. $N_9\sigma_{a9} + N_{FP}\sigma_{aFP}$ was taken into account for calculation.**

$$F(\Psi) \equiv (v\Sigma_f - \Sigma_a) = v_9 N_9 \sigma_{f9} - N_8 \sigma_{a8} - N_9 \sigma_{a9} - N_{FP}\sigma_{aFP} \quad (A2)$$

$$= N_{8,0}\left[\left(v_9 \frac{\sigma_{c8}\sigma_{f9}}{\sigma_{a9}-\sigma_{a8}} - \sigma_{a8}\right)e^{-\sigma_{a8}\Psi} - v_9 \frac{\sigma_{c8}\sigma_{f9}}{\sigma_{a9}-\sigma_{a8}}e^{-\sigma_{a9}\Psi}\right]$$

$$-N_{8,0}\frac{\sigma_{c8}}{\sigma_{a9}-\sigma_{a8}}\sigma_{a9}[e^{-\sigma_{a8}\Psi} - e^{-\sigma_{a9}\Psi}] - N_{8,0}\frac{\sigma_{c8}}{\sigma_{a9}-\sigma_{a8}}[e^{-\sigma_{a8}\Psi} - e^{-\sigma_{a9}\Psi}]\sigma_{f9}\Psi\sigma_{aFP}$$

$$= N_{8,0}\left[\left(v_9\frac{\sigma_{c8}\sigma_{f9}}{\sigma_{a9}-\sigma_{a8}}-\sigma_{a8}-\frac{\sigma_{c8}}{\sigma_{a9}-\sigma_{a8}}\sigma_{a9}-\frac{\sigma_{c8}}{\sigma_{a9}-\sigma_{a8}}\sigma_{f9}\Psi\sigma_{aFP}\right)e^{-\sigma_{a8}\Psi}-\left(v_9\frac{\sigma_{c8}\sigma_{f9}}{\sigma_{a9}-\sigma_{a8}}-\frac{\sigma_{c8}}{\sigma_{a9}-\sigma_{a8}}\sigma_{a9}-\frac{\sigma_{c8}}{\sigma_{a9}-\sigma_{a8}}\sigma_{f9}\Psi\sigma_{aFP}\right)e^{-\sigma_{a9}\Psi}\right]$$

$$= N_{8,0}\left[\frac{\sigma_{c8}}{\sigma_{a9}-\sigma_{a8}}\left(v_9\sigma_{f9}-\frac{\sigma_{a9}-\sigma_{a8}}{\sigma_{c8}}\sigma_{a8}-\sigma_{a9}\right)e^{-\sigma_{a8}\Psi}-\frac{\sigma_{c8}}{\sigma_{a9}-\sigma_{a8}}(v_9\sigma_{f9}-\sigma_{a9})e^{-\sigma_{a9}\Psi}\right]-N_{8,0}\frac{\sigma_{c8}}{\sigma_{a9}-\sigma_{a8}}\sigma_{f9}\sigma_{aFP}\Psi(e^{-\sigma_{a8}\Psi}-e^{-\sigma_{a9}\Psi})$$

Where $F(\Psi)$, $A$, $B$, $C_{FP}$, $C_8$, $C_9$ are defined as following:

$$F(\Psi) = N_{8,0}(C_8 A - C_9 B - C_{fp}\Psi(1-\sigma_{a8}\Psi) + C_{fp}\Psi(1-\sigma_{a9}\Psi)) \tag{A3}$$

$$A = 1 - \sigma_{a8}\Psi + \frac{1}{2}(\sigma_{a8}\Psi)^2$$

$$B = 1 - \sigma_{a9}\Psi + \frac{1}{2}(\sigma_{a9}\Psi)^2$$

$$C_{FP} = \frac{\sigma_{c8}}{\sigma_{a9}-\sigma_{a8}}\sigma_{f9}\sigma_{aFP} \tag{A4}$$

$$C_8 = \frac{\sigma_{c8}}{\sigma_{a9}-\sigma_{a8}}\left(v_9\sigma_{f9}-\frac{\sigma_{a9}-\sigma_{a8}}{\sigma_{c8}}\sigma_{a8}-\sigma_{a9}\right) \tag{A5}$$

$$C_9 = \frac{\sigma_{c8}}{\sigma_{a9}-\sigma_{a8}}(v_9\sigma_{f9}-\sigma_{a9}) \tag{A6}$$

$$D\frac{\partial^2 \Phi}{\partial x^2} + F(\Psi)\Phi = \frac{1}{v}\frac{\partial \Phi}{\partial t} \tag{A7}$$

Substitute $F(\Psi)$ into Eq.(A7) and apply the tanh-method,

$$2c^2 a_2 - \frac{a_2 C_8 N_{8,0}}{D} + \frac{a_2 C_9 N_{8,0}}{D} - \frac{a_2^2 C_8 \sigma_{a8} N_{8,0}}{cDu} - \frac{a_2^3 C_{fp}\sigma_{a8}N_{8,0}}{c^2 Du^2} - \frac{a_2^3 C_8 \sigma_{a8}^2 N_{8,0}}{2c^2 Du^2} + \frac{a_2^2 C_9 \sigma_{a9} N_{8,0}}{cDu} + \frac{a_2^3 C_{fp}\sigma_{a9} N_{8,0}}{c^2 Du^2} + \frac{a_2^3 C_9 \sigma_{a9}^2 N_{8,0}}{2c^2 Du^2} +$$

$$T\left(\frac{2cua_2}{Dv} + \frac{a_2^2 C_8 \sigma_{a8} N_{8,0}}{cDu} + \frac{2a_2^3 C_{fp}\sigma_{a8}N_{8,0}}{c^2 Du^2} + \frac{a_2^3 C_8 \sigma_{a8}^2 N_{8,0}}{c^2 Du^2} - \frac{a_2^2 C_9 \sigma_{a9} N_{8,0}}{cDu} - \frac{2a_2^3 C_{fp}\sigma_{a9}N_{8,0}}{c^2 Du^2} - \frac{a_2^3 C_9 \sigma_{a9}^2 N_{8,0}}{c^2 Du^2}\right) +$$

$$T^2\left(-8c^2 a_2 + \frac{a_2 C_8 N_{8,0}}{D} - \frac{a_2 C_9 N_{8,0}}{D} + \frac{a_2^2 C_8 \sigma_{a8} N_{8,0}}{cDu} - \frac{a_2^2 C_9 \sigma_{a9} N_{8,0}}{cDu}\right) +$$

$$T^3\left(-\frac{2cua_2}{Dv} - \frac{a_2^2 C_8 \sigma_{a8} N_{8,0}}{cDu} - \frac{2a_2^3 C_{fp}\sigma_{a8} N_{8,0}}{c^2 Du^2} - \frac{a_2^3 C_8 \sigma_{a8}^2 N_{8,0}}{c^2 Du^2} + \frac{a_2^2 C_9 \sigma_{a9} N_{8,0}}{cDu} + \frac{2a_2^3 C_{fp}\sigma_{a9} N_{8,0}}{c^2 Du^2} + \frac{a_2^3 C_9 \sigma_{a9}^2 N_{8,0}}{c^2 Du^2}\right) +$$

$$T^4\left(6c^2 a_2 + \frac{a_2^3 C_{fp}\sigma_{a8} N_{8,0}}{c^2 Du^2} + \frac{a_2^3 C_8 \sigma_{a8}^2 N_{8,0}}{2c^2 Du^2} - \frac{a_2^3 C_{fp}\sigma_{a9} N_{8,0}}{c^2 Du^2} - \frac{a_2^3 C_9 \sigma_{a9}^2 N_{8,0}}{2c^2 Du^2}\right) == 0$$

Collecting again all terms with the same power $T^j$ (j=0,1,2,3,4) and performing some algebra, the solution can be obtained:

$$a_2 = -\frac{2\sqrt{3D}c^2 u}{\sqrt{N_{8,0}}\sqrt{-C_8 \sigma_{a8}^2 + C_9 \sigma_{a9}^2 - 2C_{fp}\sigma_{a8} + 2C_{fp}\sigma_{a9}}} \tag{A8}$$